\documentclass[10pt,reprint,amsmath,amssymb,aps,pra,showpacs,longbibliography,superscriptaddress]{revtex4-1}
\usepackage[latin9]{inputenc}
\usepackage{amsmath}
\usepackage{amssymb}
\usepackage{graphicx}
\usepackage{esint}

\makeatletter

\usepackage{amsfonts}
\usepackage{graphics}

\makeatother

\begin{document}
\title{Super Fermi polaron and Nagaoka ferromagnetism in a two-dimesnional
square lattice}
\author{Hui Hu}
\email{hhu@swin.edu.au}

\affiliation{Centre for Quantum Technology Theory, Swinburne University of Technology,
Melbourne 3122, Australia}
\author{Jia Wang}
\email{jiawang@swin.edu.au}

\affiliation{Centre for Quantum Technology Theory, Swinburne University of Technology,
Melbourne 3122, Australia}
\author{Xia-Ji Liu}
\email{xiajiliu@swin.edu.au}

\affiliation{Centre for Quantum Technology Theory, Swinburne University of Technology,
Melbourne 3122, Australia}
\date{\today}
\begin{abstract}
We consider the Fermi polaron problem of an impurity hopping around
a two-dimensional square lattice and interacting with a sea of fermions
at given filling factor. When the interaction is attractive, we find
standard Fermi polaron quasiparticles, categorized as attractive polarons
and repulsive polarons. When the interaction becomes repulsive, interestingly,
we observe an unconventional highly-excited polaron quasiparticle,
sharply peaked at the corner of the first Brillouin zone with momentum
$\mathbf{k}=(\pm\pi,\pm\pi)$. This super Fermi polaron branch arises
from the dressing of the impurity's motion with holes, instead of
particles of fermions. We show that super Fermi polarons become increasingly
well-defined with increasing impurity-fermion repulsions and might
be considered as a precursor of Nagaoka ferromagnetism, which would
appear at sufficiently large repulsions and at large filling factors.
We also investigate the temperature-dependence of super Fermi polarons
and find that they are thermally robust against the significant increase
in temperature.
\end{abstract}
\maketitle

\section{Introduction}

The problem of a single impurity moving in a many-body environment
of a non-interacting Fermi sea is probably the simplest quantum many-body
system \cite{Alexandrov2010}. It was first addressed by Lev Landau
ninety years ago in a two-page short paper \cite{Landau1933}, which
gives birth to a fundamental concept known as quasiparticle. Termed
as a ``polaron'' quasiparticle, or more specifically ``Fermi polaron''
to reflect the Fermi sea background, this impurity problem arises
in diverse research fields, including Kondo screening \cite{Hewson1993,Zhang2001,Snyman2023},
Anderson\textquoteright s orthogonality catastrophe \cite{Anderson1967},
the X-ray Fermi edge singularity \cite{Mahan1967,Roulet1969,Nozieres1969},
Nagaoka ferromagnetism \cite{Nagaoka1966,Shastry1990,Basile1990,vonderLinden1991,Cui2010},
the phase string effect \cite{Sheng1996}, ultracold atomic polaron
\cite{Chevy2006,Schirotzek2009,Zhang2012,Massignan2014,Scazza2017,Schmidt2018,Wang2022PRL,Wang2022PRA,Wang2023AB},
and most recently exciton-polariton polaron in two-dimensional materials
\cite{Sidler2017,Hu2023AB}.

Among those fields, the recent research on ultracold atomic polarons
attracts particular interests, due to the unprecedented tunability
and controllability on quantum atomic gases \cite{Bloch2008}. For
instance, the interatomic interaction can be precisely tuned by changing
an external magnetic field across a Feshbach resonance \cite{Chin2010}.
In the strongly interacting regime near resonance, particle-hole excitations
of the Fermi sea are attached to the impurity \cite{Chevy2006}, forming
an attractive Fermi polaron in the absolute ground state. Moreover,
above the Feshbach resonance in the presence of a two-body bound (molecule)
state, although the underlying interaction between the impurity and
the Fermi sea is always attractive, an additional repulsive Fermi
polaron develops as an excited state \cite{Cui2010,Massignan2011}.
Over the past fifteen years, attractive and repulsive Fermi polarons
have been extensively investigated in a quantitative manner, both
experimentally \cite{Schirotzek2009,Zhang2012,Scazza2017,Zan2019,Cetina2016,Ness2020}
and theoretically \cite{Chevy2006,Wang2022PRL,Wang2022PRA,Wang2023AB,Combescot2007,Prokofev2008,Parish2011,Knap2012,Parish2013,Vlietinck2013,Bour2015,Goulko2016,Hu2018,Tajima2018,Liu2019,Wang2019,Tajima2019,Hu2022a,Hu2022b,Hu2022c,Hu2023ABpwave}. 

In this work, we would like to suggest the existence of a novel Fermi
polaron in two-dimensional (2D) optical lattices, which can be readily
realized in cold-atom experiments, again owing to the unprecedented
tunability and controllability. In lattices, the interaction between
the impurity and fermions in the Fermi sea can become positive. In
addition, the filling factor $\nu$ of fermions can be tuned to be
near unity ($\nu\sim1$), so the background environment can be more
conveniently described as a new Fermi sea of holes, centered around
the corner of the first Brillouin zone, where $\mathbf{k}=(k_{x},k_{y})=(\pm\pi,\pm\pi)$.
Therefore, the repulsion between the impurity and fermions can be
equivalently treated as an effective attraction between the impurity
and holes, leading to an ``attractive'' Fermi polaron that has the
\emph{highest} energy. This state is analogous to the highly excited
super-Tonk-Girardeau gas phase found in a one-dimensional Bose gas
with infinitely strong attraction \cite{Astrakharchik2005}, which
is contrasted with a ground-state Tonk-Girardeau gas with infinitely
strong repulsions. Thus, it is useful to dub the novel highest-lying
Fermi polaron as a \emph{super} Fermi polaron.

We find that the appearance of the super Fermi polaron is a precursor
of Nagaoka ferromagnetism \cite{Nagaoka1966}, which is anticipated
to occur at large repulsion and at large filling factor for a cluster
of spin-1/2 fermions \cite{Shastry1990,Basile1990,vonderLinden1991,Cui2010}.
There, all fermions prefer to align their spin (i.e., into the spin-up
state), due to the strong repulsion between two fermions with unlike
spin. In other words, if initially we consider a spin-down fermion
(i.e., impurity) immersed in a sea of spin-up fermions, the spin-down
fermion at zero momentum prefers to flip its spin, occupies into a
spin-up state near Fermi surface located at about $\mathbf{k}=(\pm\pi,\pm\pi)$,
and eventually creates a fully spin-polarized non-interacting Fermi
sea. In our case of super Fermi polaron, of course, the impurity can
not take spin-flip and automatically turn itself into a fermion. However,
this tendency is clearly demonstrated in the impurity spectral function:
on the one hand, near zero momentum the quasiparticle peak becomes
increasingly blurred in the spectral function; on the other hand,
a very sharp peak well develops at $\mathbf{k}=(\pm\pi,\pm\pi)$.
To further confirm the instability towards Nagaoka ferromagnetism,
at low temperature we calculate the ground-state energy of the super
Fermi polaron and find that it is indeed preferable in energy to take
an imaginable ``spin-flip'' at large repulsion. 

The rest of the paper is laid out as follows. In the next section
(Sec. II), we describe the model Hamiltonian for an impurity interacting
with a Fermi sea on a 2D square lattice with on-site interaction.
In Sec. III, we solve the model Hamiltonian at finite temperature,
by using a non-self-consistent many-body $T$-matrix approach that
captures the crucial one-particle-hole excitations of the Fermi sea.
In Sec. IV, we first report the results of conventional Fermi polarons
with an attractive on-site interaction strength $U<0$. We then consider
a repulsive on-site interaction ($U>0$) and discuss the evolution
of the impurity spectral function as functions of the filling factor
and repulsion strength. We clearly demonstrate the Nagaoka ferromagnetic
transition by comparing the energies of the Fermi polaron state and
of the fully polarized Fermi sea, and determine the critical interaction
strength at a given filling factor. We finally discuss the temperature
dependence of Fermi polarons, and show the remarkable thermal robustness
of the super Fermi polaron. We conclude in Sec. V and present an outlook
for future studies. 

\section{The model Hamiltonian}

Let us start by considering one impurity and $N$ fermionic atoms
moving on a 2D $L\times L$ square optical lattice, with hopping strengths
$t_{d}$ and $t$, resepctively. The impurity interacts with fermions
when they occupy the same site only. In momentum space, the system
can be described by the standard Hubbard model,

\begin{equation}
\mathcal{H}=\sum_{\mathbf{k}}\left(\xi_{\mathbf{k}}c_{\mathbf{k}}^{\dagger}c_{\mathbf{k}}+E_{\mathbf{k}}d_{\mathbf{k}}^{\dagger}d_{\mathbf{k}}\right)+\frac{U}{A}\sum_{\mathbf{kk'q}}c_{\mathbf{k}}^{\dagger}d_{\mathbf{q}-\mathbf{k}}^{\dagger}d_{\mathbf{q}-\mathbf{k'}}c_{\mathbf{k}'},\label{eq:ModelHamiltonian}
\end{equation}
where $A=(La)^{2}$ is the area of the system with a lattice spacing
$a$, $c_{\mathbf{k}}^{\dagger}$ and $d_{\mathbf{k}}^{\dagger}$
are the creation field operators for fermionic atoms and the impurity,
respectively. The first term of the model Hamiltonian describes the
single-particle motion with dispersion relation 
\begin{equation}
\xi_{\mathbf{k}}=-2t\left(\cos k_{x}+\cos k_{y}\right)-\mu
\end{equation}
for atoms and 
\begin{equation}
E_{\mathbf{k}}=-2t_{d}\left(\cos k_{x}+\cos k_{y}\right)
\end{equation}
for the impurity, while the last term is the interaction Hamiltonian
with on-site interaction strength $U$. Here, for convenience we have
taken the lattice size $a=1$, so the first Brillouin zone is given
by $k_{x},k_{y}\subseteq[-\pi,+\pi]$. We have introduced a chemical
potential $\mu$ to tune the filling factor $\nu=N/A$ of atoms on
the lattice. In the thermodynamic limit (i.e., $N\rightarrow\infty$),
the motion of fermionic atoms is barely affected by the existence
of the impurity, so at finite temperature $T$, the chemical $\mu$
simply relates to $\nu$ by the non-interacting number equation,
\begin{equation}
\nu=\frac{1}{A}\sum_{\mathbf{k}}\left\langle c_{\mathbf{k}}^{\dagger}c_{\mathbf{k}}\right\rangle =\intop_{-\pi}^{+\pi}\frac{dk_{x}dk_{y}}{\left(2\pi\right)^{2}}f\left(\xi_{\mathbf{k}}\right),
\end{equation}
where $f(x)=1/[e^{x/(k_{B}T)}+1]$ is the Fermi-Dirac distribution
function. In contrast, the behavior of the impurity is strongly modified
by the on-site interaction and could be solved below by using a non-self-consistent
many-body $T$-matrix theory. Moreover, for a single impurity, it
is not necessary to explicitly introduce an impurity chemical potential
\cite{Hu2022a}. Throughout the work, we always assume the impurity
and fermionic atoms have the same hopping strength, i.e., $t_{d}=t$,
and we use $t$ as the units of energy.

\section{Non-self-consistent many-body $T$-matrix approach}

The non-self-consistent many-body $T$-matrix theory of Fermi polarons
has been thoroughly studied in the past, without considering optical
lattices. The generalization of the theory to the lattice case is
straightforward, since the resulting equations for the key quantities,
such as the vertex function and the impurity self-energy, take the
exactly same forms. The only change is to restrict the summation over
the momentum $\mathbf{k}=(k_{x},k_{y})$ to the first Brillouin zone.
Therefore, we directly write down the inverse vertex function \cite{Combescot2007,Hu2022a},
\begin{equation}
\Gamma^{-1}\left(\mathbf{q},\omega\right)=\frac{1}{U}-\intop_{-\pi}^{+\pi}\frac{dk_{x}dk_{y}}{\left(2\pi\right)^{2}}\frac{1-f\left(\xi_{\mathbf{k}}\right)}{\omega-\xi_{\mathbf{k}}-E_{\mathbf{q}-\mathbf{k}}},\label{eq:VertexFunction}
\end{equation}
and the impurity self-energy,
\begin{equation}
\Sigma\left(\mathbf{k},\omega\right)=\intop_{-\pi}^{+\pi}\frac{dq_{x}dq_{y}}{\left(2\pi\right)^{2}}f\left(\xi_{\mathbf{q}-\mathbf{k}}\right)\Gamma\left(\mathbf{q},\omega+\xi_{\mathbf{q}-\mathbf{k}}\right).\label{eq:SelfEnergy}
\end{equation}
Once the impurity self-energy is determined, we calculate the impurity
Green function \cite{Combescot2007,Hu2022a},
\begin{equation}
G\left(\mathbf{k},\omega\right)=\frac{1}{\omega-E_{\mathbf{k}}-\Sigma\left(\mathbf{k},\omega\right)}.
\end{equation}
The emergent Fermi polarons can be well-characterized by the impurity
spectral function 
\begin{equation}
A\left(\mathbf{k},\omega\right)=-\frac{1}{\pi}\textrm{Im}G\left(\mathbf{k},\omega\right),
\end{equation}
where the existence of polaron quasiparticles is clearly revealed
by a sharp spectral peak. The position and the width of the spectral
peak relate to the energy $\mathcal{E}_{P}(\mathbf{k})$ and the decay
rate (i.e., inverse lifetime) $\Gamma_{P}(\mathbf{k})$ of polaron
quasiparticles \cite{Massignan2011,Combescot2007}, respectively.
It is readily seen that the polaron energy $\mathcal{E}_{P}(\mathbf{k})$,
of either attractive Fermi polaron or repulsive Fermi polaron, is
given by the pole of the impurity Green function, if we take the replacement
$\omega\rightarrow\mathcal{E}_{P}(\mathbf{k)}$: 
\begin{equation}
\mathcal{E}_{P}\left(\mathbf{k}\right)=E_{\mathbf{k}}+\textrm{Re}\Sigma\left[\mathbf{k},\mathcal{E}_{P}\left(\mathbf{k}\right)\right].\label{eq:PolaronEnergy}
\end{equation}
By Taylor-expanding the impurity self-energy around the polaron energy,
i.e., 
\begin{equation}
\Sigma\left(\omega\right)\simeq\textrm{Re}\Sigma\left(\mathcal{E}_{P}\right)+\frac{\partial\textrm{Re}\Sigma(\omega)}{\partial\omega}\left(\omega-\mathcal{E}_{P}\right)+i\textrm{Im}\Sigma\left(\mathcal{E}_{P}\right),
\end{equation}
where we have suppressed the dependence of the impurity self-energy
on the momentum $\mathbf{k}$, the impurity spectral function takes
an approximate Lorentzian form in the vicinity of the polaron energy,
\begin{equation}
A\left(\mathbf{k},\omega\right)\simeq\frac{\mathcal{Z}_{\mathbf{k}}}{\pi}\frac{\Gamma_{P}\left(\mathbf{k}\right)/2}{\left[\omega-\mathcal{E}_{P}\left(\mathbf{k}\right)\right]^{2}+\Gamma_{P}^{2}\left(\mathbf{k}\right)/4}.
\end{equation}
Here, $\mathcal{Z}_{\mathbf{k}}$ is the polaron residue,
\begin{equation}
\mathcal{Z}_{\mathbf{k}}=\left[1-\left.\frac{\partial\textrm{Re}\Sigma(\mathbf{k},\omega)}{\partial\omega}\right|_{\omega=\mathcal{E}_{P}\left(\mathbf{k}\right)}\right]^{-1},
\end{equation}
and $\Gamma_{P}(\mathbf{k})$ is the polaron decay rate,
\begin{equation}
\Gamma_{P}\left(\mathbf{k}\right)=-2\mathcal{Z}_{\mathbf{k}}\textrm{Im}\Sigma\left[\mathbf{k},\mathcal{E}_{P}\left(\mathbf{k}\right)\right].
\end{equation}
In the dilute limit of vanishingly small filling factor $\nu\rightarrow0$,
where the interesting physics occurs at the very small momentum, the
system behaves like an interacting Fermi gas in free space with a
contact interaction potential. In this case, for a negative on-site
interaction strength $U<0$ and an associated binding energy $\varepsilon_{B}$,
we may then introduce a dimensionless interaction parameter 
\begin{equation}
\zeta=\frac{1}{2}\ln\left(2\varepsilon_{F}/\varepsilon_{B}\right)\label{eq:eta}
\end{equation}
 to fully characterize the universal low-energy polaron physics \cite{Parish2011,Bour2015}.

It should be note that, at zero temperature our non-self-consistent
many-body $T$-matrix theory is fully equivalent to a variational
Chevy ansatz \cite{Cui2010}, which has been extensively used in the
investigations of Nagaoka ferromagnetism \cite{Shastry1990,Basile1990,vonderLinden1991},
particularly for the idealized case of infinitely large repulsion
($U\rightarrow+\infty$). Thus, our work might be viewed as a useful
extension of these variational studies to the realistic cases with
large but finite repulsion at nonzero temperature.

The key difficulty of applying the non-self-consistent $T$-matrix
theory for Fermi polarons comes from the numerical integration over
the momentum $\mathbf{k}$ in Eq. (\ref{eq:VertexFunction}). This
is caused by the singularity in the integrand, which occurs when the
energy or frequency $\omega$ lies in the two-particle scattering
continuum, i.e., $\omega=\xi_{\mathbf{k}}+E_{\mathbf{q}-\mathbf{k}}$,
at certain momenta $\mathbf{k}$. A formal procedure to solve the
difficulty is to first calculate the imaginary part of the vertex
function 
\begin{equation}
\textrm{Im}\Gamma^{-1}\left(\mathbf{q},\omega\right)=\frac{\pi}{A}\sum_{\mathbf{k}}\left[1-f\left(\xi_{\mathbf{k}}\right)\right]\delta\left(\omega-\xi_{\mathbf{k}}-E_{\mathbf{q}-\mathbf{k}}\right),
\end{equation}
where $\delta(x)$ is the Dirac delta funciton \cite{Hu2022a}. We
then use the Kramers--Kronig relation to recover the real part of
the vertex function. In our case, since the momentum is restricted
to the first Brillouin zone, a more economic and straightforward way
is to introduce a nonzero broadening factor $\eta$ and replace the
frequency $\omega$ with $\omega+i\eta$ to remove the singularity
of the integrand. We then extrapolate the results to the zero-broadening
limit. In practice, we find the following linear extrapolation with
a broadening factor $\eta=0.3t$, 
\begin{equation}
\Gamma^{-1}\left(\mathbf{q},\omega\right)\simeq2\Gamma^{-1}\left(\mathbf{q},\omega+i\eta\right)-\Gamma^{-1}\left(\mathbf{q},\omega+2i\eta\right),\label{eq:LinearExtrapolation}
\end{equation}
works very well. The choice of the value $\eta=0.3t$ is discussed
in Appendix A. 

\begin{figure}
\begin{centering}
\includegraphics[width=0.25\textwidth]{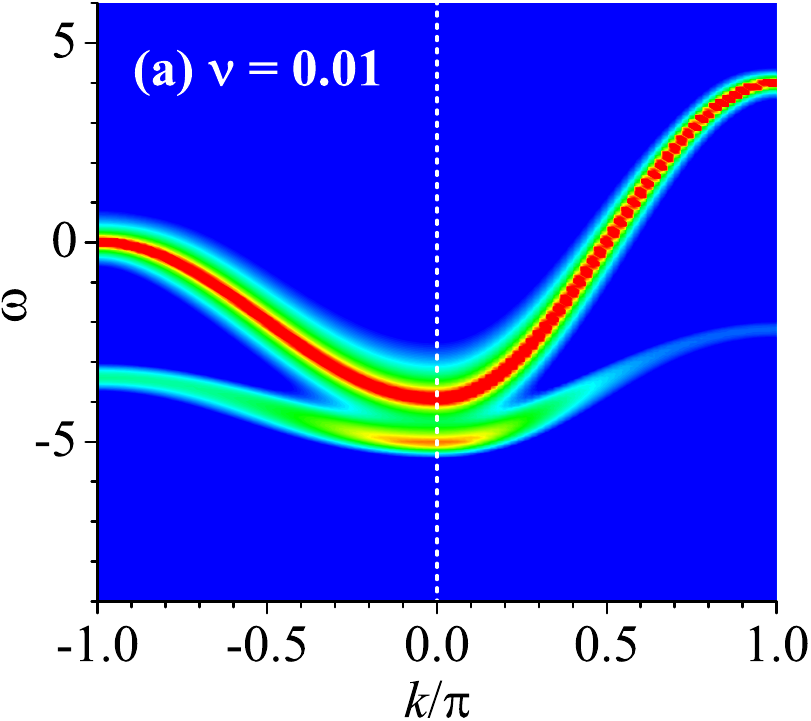}\includegraphics[width=0.25\textwidth]{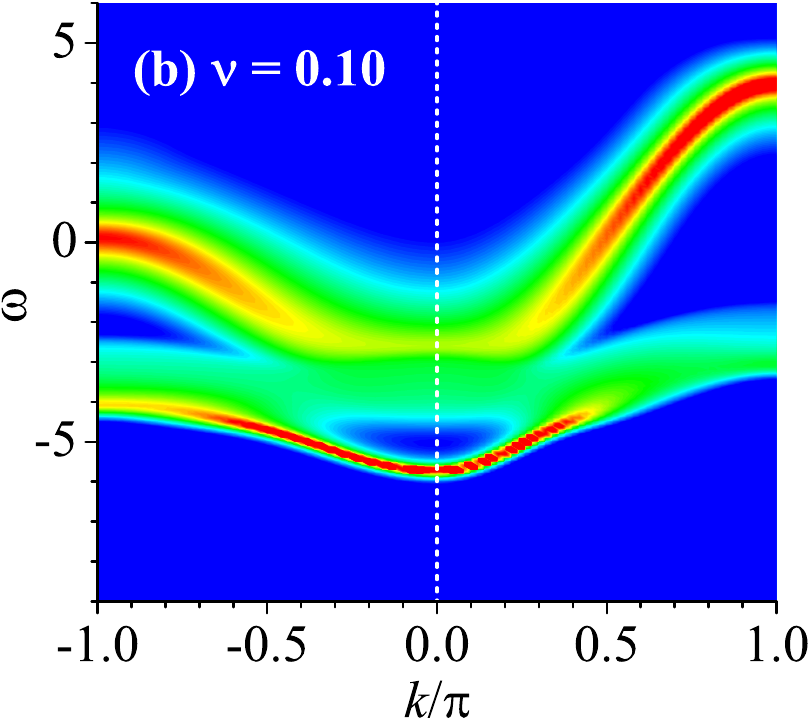}
\par\end{centering}
\begin{centering}
\includegraphics[width=0.25\textwidth]{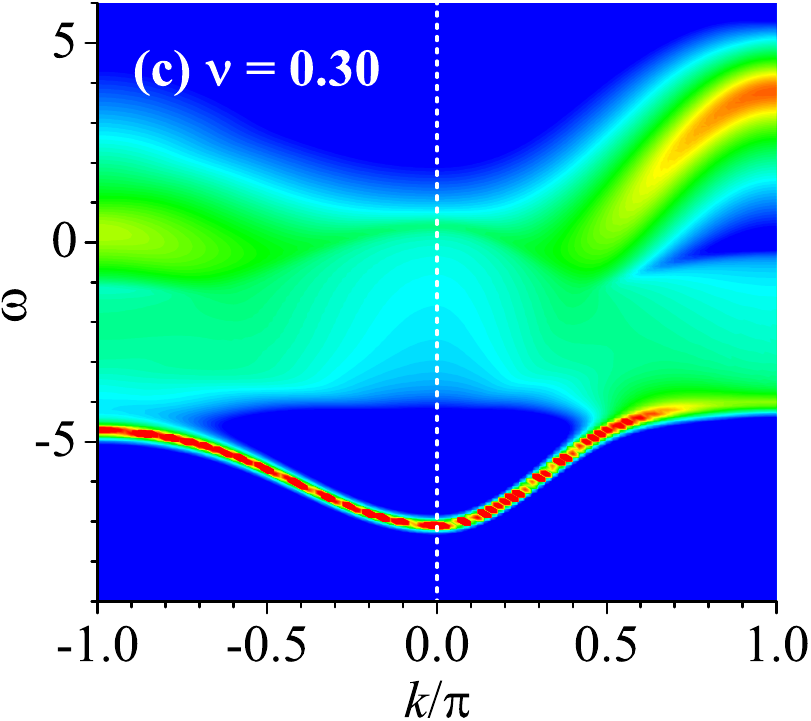}\includegraphics[width=0.25\textwidth]{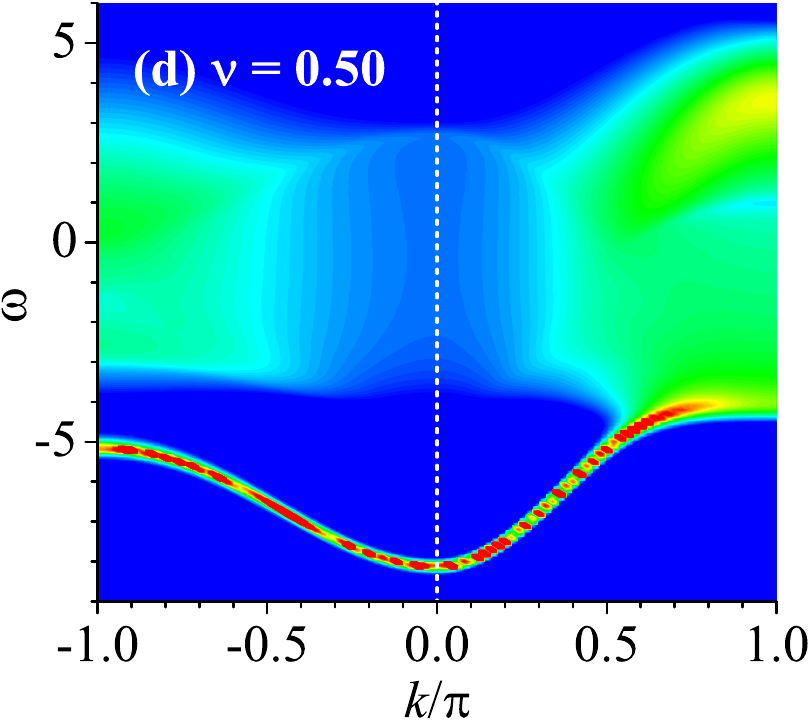}
\par\end{centering}
\caption{\label{fig:fig1_akw2dU6m} Impurity spectral function $A(\mathbf{k},\omega)$
at different filling factors $\nu$ and at a negative interaction
strength $U=-6t$, shown as 2D contour plots with a logarithmic scale
in units of $t^{-1}$. The blue and red colors represent the minimum
intensity (i.e., $0.01t^{-1}$) and maximum intensity (i.e., $t^{-1}$),
respectively. On the left hand side of each panel (i.e., $k<0$),
we consider a cut in the first Brillouin zone from the $\Gamma$-point
$\Gamma=(0,0)$ to the $X$-point $X=(\pi,0$), so $k$ represents
the wavevector $\mathbf{k}=(k,0)$. On the right hand side of the
dotted line (i.e., $k>0$), the cut is along the diagonal direction
from the $\Gamma$-point to the $M$-point $M=(\pi,\pi)$. so $k$
gives the wavevector $\mathbf{k}=(k,k)$. The energy $\omega$ is
in units of $t$ and the temperature is set to $k_{B}T=0.1t$. }
\end{figure}

\begin{figure}
\begin{centering}
\includegraphics[width=0.5\textwidth]{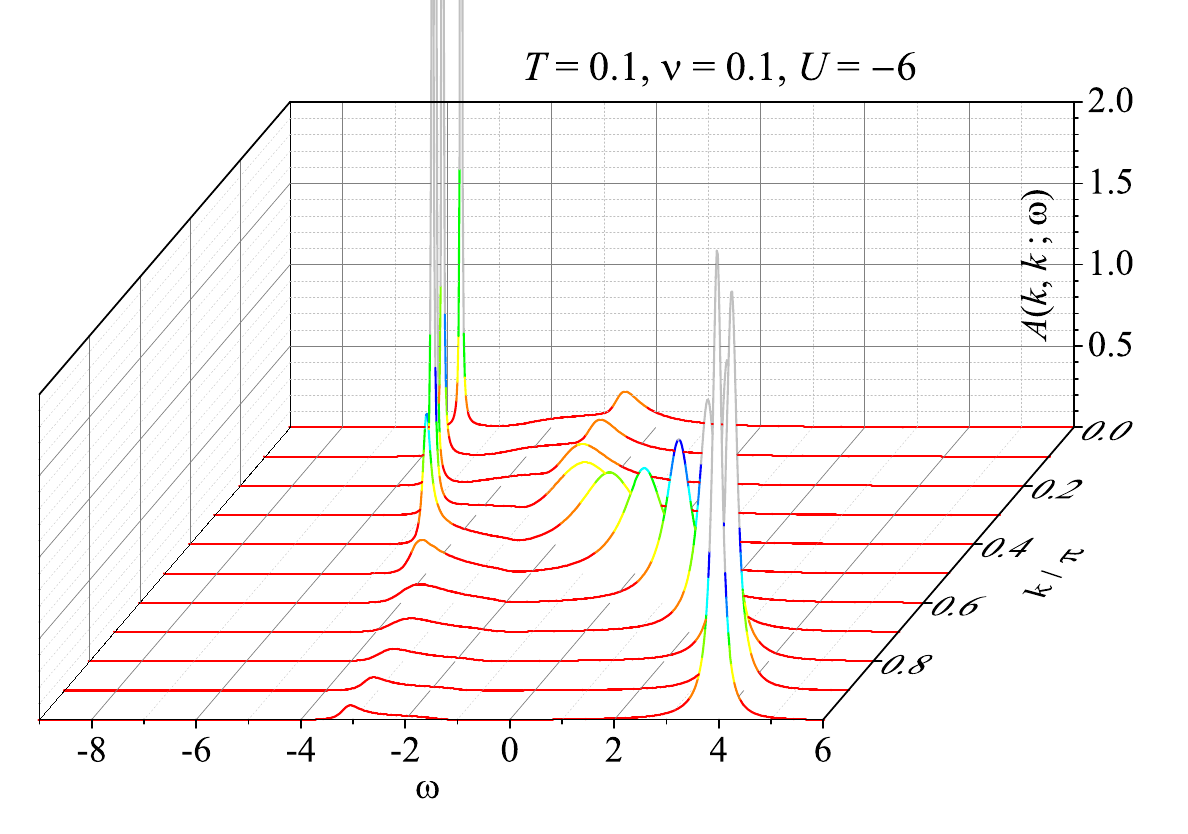}
\par\end{centering}
\caption{\label{fig:fig2_akw1dU6mNu010} Impurity spectral function $A(k_{x}=k,k_{y}=k;\omega)$
along the diagonal direction of the first Brillouin zone (i.e., the
$\Gamma M$ line), at temperature $k_{B}T=0.1t$, filling factor $\nu=0.1$
and a negative interaction strength $U=-6t$. The energy $\omega$
and the spectral function $A(\mathbf{k},\omega$) are in units of
$t$ and $t^{-1}$, respectively.}
\end{figure}

\section{Results and discussions}

\subsection{Fermi polarons at $U<0$}

Let us first consider the cases with negative on-site interaction
strengths $U<0$, which connect to the well-studied 2D Fermi polarons
in free space \cite{Massignan2014,Parish2011,Parish2013}. These cases
have also been investigated by using an ab-initio impurity lattice
Monte Carlo method \cite{Bour2015}. However, the ab-initio results
are restricted to the polaron energy only.

In Fig. \ref{fig:fig1_akw2dU6m}, we present the impurity spectral
function at $U=-6t$ and at four filling factors (as indicated), in
the form of a 2D contour plot with a logarithmic scale, where the
spectral peaks in red color can be clearly identified. To account
for the anistropy of the first Brillouin zone, in each panel, following
the convention we consider two cuts on the Brillouin zone along the
$\Gamma X$ line (see the left part) and the $\Gamma M$ line (the
right part). As the filling factor $\nu$ increases, we always find
two branches in the spectral function: the low-energy attractive Fermi
polarons and high-energy repulsive Fermi polarons. However, the evolutions
of the two kinds of Fermi polarons as a function of the filling factor
turn out be very different.

For the very-low filling factor $\nu=0.01$ in Fig. \ref{fig:fig1_akw2dU6m}(a),
the spectral function is dominated by the repulsive polaron branch.
The attractive polaron branch is only visible at low momentum $\mathbf{k}\sim0$.
Towards the $X$ point or the $M$ point, the spectral weight of attractive
Fermi polarons quickly disappears. This weak attractive polaron branch
might be understood from the results of 2D Fermi polarons in free
space. In the dilute limit at zero temperature, the dispersion relation
of fermionic atoms can be well approximated as 
\begin{equation}
\xi_{\mathbf{k}}\simeq t\mathbf{k}^{2}-(\mu+4t)=\frac{\hbar^{2}\mathbf{k}^{2}}{2m}-\varepsilon_{F},
\end{equation}
where $m=\hbar^{2}/(2t)$ is the effective mass and $\varepsilon_{F}=\hbar^{2}k_{F}^{2}/(2m)$
is the Fermi energy with Fermi wavevector $k_{F}=(4\pi\nu)^{1/2}$.
It is easy to see that $\varepsilon_{F}=4\pi\nu t$. At $U=-6t$,
the binding energy $\varepsilon_{B}\sim t$. Thus, the dimensionless
interaction parameter in Eq. (\ref{eq:eta}) is about $\zeta\sim(1/2)\ln(8\pi\nu)\simeq-0.7$,
which is very close to the threshold for the polaron-molecule transition
\cite{Parish2011,Bour2015}. At this interaction parameter, the residue
$\mathcal{Z}_{\mathbf{k}\sim0}$ for the attractive polaron is not
significant. 

\begin{figure}
\begin{centering}
\includegraphics[width=0.5\textwidth]{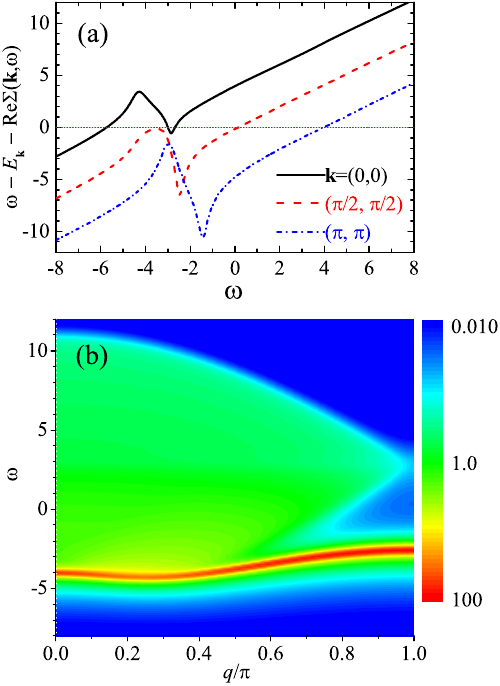}
\par\end{centering}
\caption{\label{fig:fig3_segkwU6mNu010} (a) The real part of the inverse of
the impurity Green function $\textrm{Re}G^{-1}(\mathbf{k},\omega)$
as a function of the energy $\omega$, at three different wave-vectors:
the $\Gamma$-point (black solid line), $\mathbf{k}=(\pi/2,\pi/2)$
(red dashed line), and the $M$-point (blue dot-dashed line). The
pole of the impurity Green function, at which $\omega-E_{\mathbf{k}}-\Sigma(\mathbf{k},\omega)=0$
as given by the crossing point with the green dotted line, deterimines
the energy of polaron quasiparticles. (b) The molecular spectral function
$A_{\textrm{mol}}(q_{x}=q,q_{y}=q;\omega)$ along the diagonal direction
in momentum space, in arbitrary units (as indicated by the color bar
in the logarithmic scale). Here, we take the same parameters for $T$,
$\nu$ and $U$ as in Fig. \ref{fig:fig2_akw1dU6mNu010}.}
\end{figure}

\begin{figure*}
\begin{centering}
\includegraphics[width=0.33\textwidth]{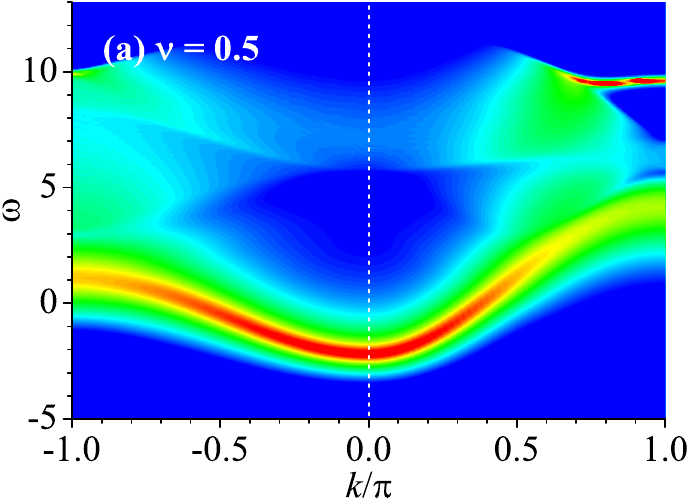}\includegraphics[width=0.33\textwidth]{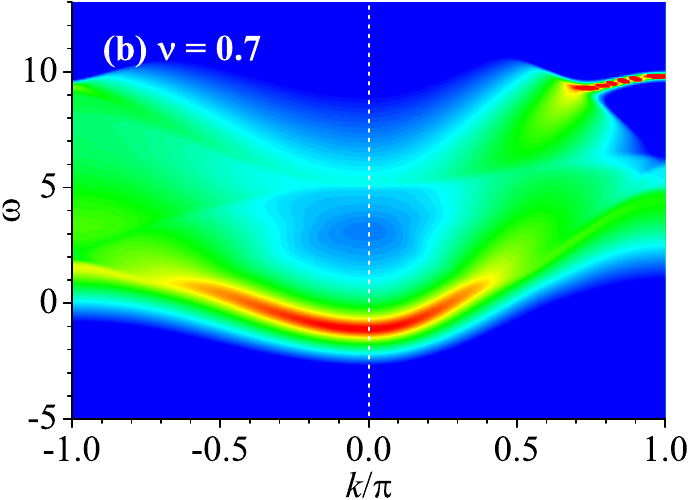}\includegraphics[width=0.33\textwidth]{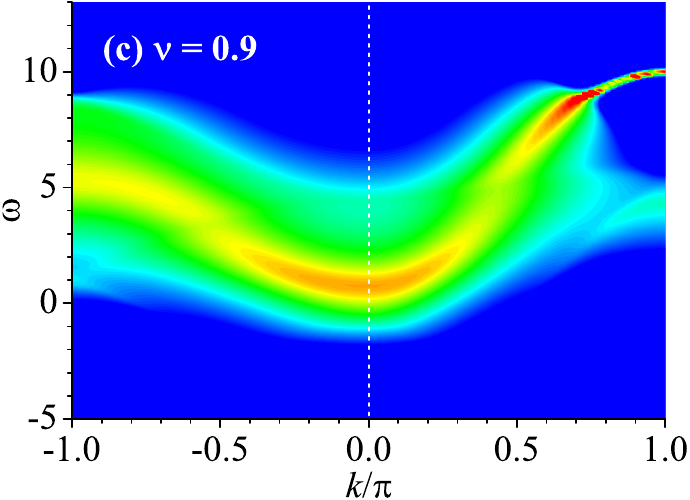}
\par\end{centering}
\caption{\label{fig:fig4_akw2dU6p} Impurity spectral function $A(\mathbf{k},\omega)$
at three filling factors $\nu=0.5$ (a), $\nu=0.7$ (b) and $\nu=0.9$
(c), and at a positive interaction strength $U=6t$. The same logarithmic
contour plots as in Fig. \ref{fig:fig1_akw2dU6m} have been used.
The temperature is set to $k_{B}T=0.1t$ as well.}
\end{figure*}

For the low filling factor $\nu=0.1$ in Fig. \ref{fig:fig1_akw2dU6m}(b),
the dimensionless interaction strength $\zeta$ increases to about
$\zeta\sim0.5$, where near zero momentum the attractive Fermi polaron
is well defined. Indeed, we find that a much sharper attractive polaron
peak with large spectral weight or polaron residue $\mathcal{Z}_{\mathbf{k}}$.
Accordingly, the repulsive polaron peak near zero momentum becomes
blurred, with much smaller spectral weight. Interestingly, the repulsive
Fermi polaron at large momentum near the $X$ point or the $M$ point
remains sharply peaked. This observation is in marked contrast to
the free space 2D polaron model, where Fermi polarons always become
less well-defined at large momentum. Therefore, this feature should
be related to the unique structure of square lattice. We may understand
it as a consequence of the Van Hove singularity is in the density
of states of square lattice. In particular, the logarithmically divergent
density of states at the $X$ point could be energetically favorable
for particle-hole excitations and therefore leads to more stable Fermi
polarons. On the other hand, a well-marked repulsive polaron at the
$M$ point seems to indicate a more robust two-body bound state at
the corner Brillouin zone than at zero momentum. To better show the
robust repulsive Fermi polaron, we also report in Fig. \ref{fig:fig2_akw1dU6mNu010}
the evolution of the one-dimensional spectral function as the momentum
increases along the $\Gamma M$ line.

As we further increase the filling factor $\nu$, the effect the square
lattice band structure becomes more prominent. As shown in Fig. \ref{fig:fig1_akw2dU6m}(c)
and Fig. \ref{fig:fig1_akw2dU6m}(d) for $\nu=0.3$ and $\nu=0.5$,
the spectrum of the upper repulsive Fermi polaron distributes much
wider, in sharp contrast to the attractive Fermi polaron, whose spectral
response becomes increasingly narrower. Nevertheless, even at $\nu=0.5$
the repulsive polaron peak near the $M$ point remains visible, although
the spectral weight of the repulsive branch gets strongly depleted
close to the $\Gamma$ point at zero momentum.

To better understand the robust repulsive Fermi polaron near the $M$
point, we focus on the case $\nu=0.1$ and report in Fig. \ref{fig:fig3_segkwU6mNu010}(a)
the real part of the inverse impurity green function 
\begin{equation}
\textrm{Re}G^{-1}\left(\mathbf{k},\omega\right)=\omega-E_{\mathbf{k}}-\textrm{Re}\Sigma\left(\mathbf{k},\omega\right)
\end{equation}
at three different momenta. The condition of a pole in the impurity
green function, i.e., $\textrm{Re}G^{-1}(\mathbf{k},\omega)=0$, determines
the polaron energy. We observe that although in general $\textrm{Re}G^{-1}(\mathbf{k},\omega)$
increases with the frequency $\omega$, it has a peculiar peak-dip
structure around the bottom of the energy band (i.e., $\omega\sim-4t$
at zero momentum). The depth of this peak-dip structure increases
with increasing momentum. At $\mathbf{k}=0$, we find two solutions
of $\textrm{Re}G^{-1}(\mathbf{k},\omega)=0$, giving rise to an attractive
Fermi polaron at $\omega\simeq-6t$ and a repulsive Fermi polaron
at $\omega\simeq-3t$. At $\mathbf{k}=(\pi/2,\pi/2)$ and at the $M$
point with $\mathbf{k}_{M}=(\pi,\pi)$ , the peak values of $\textrm{Re}G^{-1}(\mathbf{k},\omega)$
become negative, implying the absence of the attractive polaron. However,
the repulsive Fermi polaron always appears, owing to the positive
and $k$-independent slope with increasing frequency. 

The existence of repulsive Fermi polaron is generally related to a
two-body molecule bound state. In Fig. \ref{fig:fig3_segkwU6mNu010}(b),
we present the molecule spectral function 
\begin{equation}
A_{\textrm{mol}}\left(\mathbf{q},\omega\right)=-\frac{1}{\pi}\textrm{Im}\Gamma\left(\mathbf{q},\omega\right)
\end{equation}
along the $\Gamma M$ line in the form of a 2D contour plot. We may
clearly identify the two-particle scattering continuum enclosed by
$\omega_{\textrm{min}}(\mathbf{q})=\min_{\{\mathbf{k}\}}(\xi_{\mathbf{k}}+E_{\mathbf{q}-\mathbf{k}})$
and $\omega_{\textrm{max}}(\mathbf{q})=\max_{\{\mathbf{k}\}}(\xi_{\mathbf{k}}+E_{\mathbf{q}-\mathbf{k}})$.
There is always an in-medium molecule bound state with energy $\mathcal{E}_{M}(\mathbf{q})$,
as revealed by a strong spectral peak near the bottom of the scattering
continuum. The dispersion of $\mathcal{E}_{M}(\mathbf{q})$ is non-monotonic
and exhibits a minimum at about $q\sim0.3\pi$. More importantly,
at $q<q_{c}\sim\pi/2$, the molecule state is buried in the scattering
continuum, so the molecule peak has a finite spectral width due to
scattering and can be viewed as quasi-bound state. Above $q_{c}$,
the molecule state develops into a true long-lived bound state, although
there is a residual spectral width due to thermal broadening. Near
the $M$ point, therefore the molecule state becomes very robust.
This robustness directly leads to the well-defined repulsive Fermi
polaron at the corner of the first Brillouin zone, as we highlighted
earlier.

\begin{figure}
\begin{centering}
\includegraphics[width=0.45\textwidth]{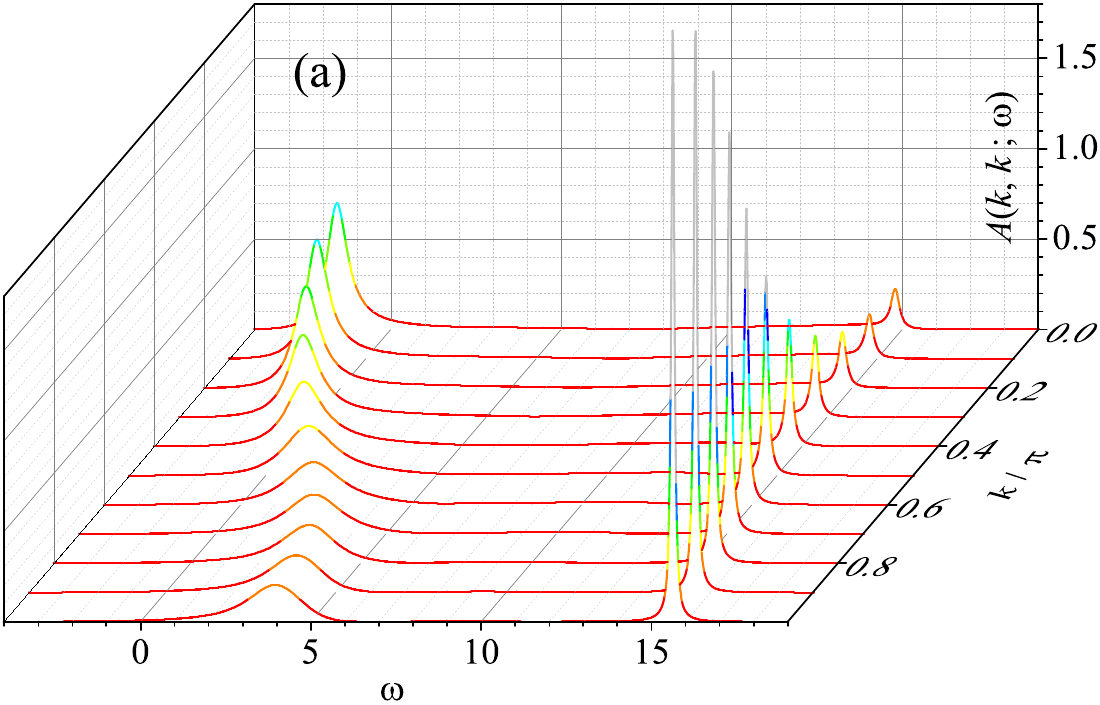}
\par\end{centering}
\begin{centering}
\includegraphics[width=0.45\textwidth]{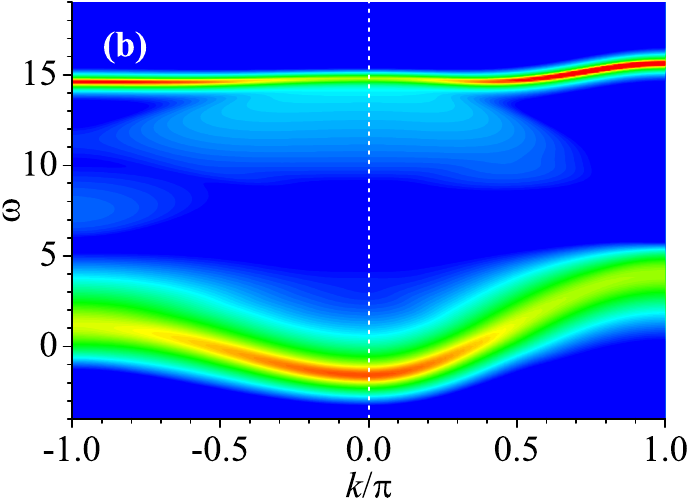}
\par\end{centering}
\caption{\label{fig:fig5_akw2dU12p} Impurity spectral function $A(\mathbf{k},\omega)$
at a large positive interaction strength $U=12t$, where the upper
branch of super Fermi polarons becomes well-defined for all wave-vectors.
We show the spectral function in either the standard one-dimensional
plot along the diagonal direction $k_{x}=k_{y}=k$ (a) or in the 2D
contour plot with a logarithmic scale (b). The parameters $k_{B}T=0.1t$
and $\nu=0.5$ have been used.}
\end{figure}

\subsection{Super Fermi polarons at $U>0$}

We now turn to consider a positive on-site interaction $U>0$. In
Fig. \ref{fig:fig4_akw2dU6p}, we show the 2D contour plots of spectral
function at $U=+6t$ and at three filling factors, $\nu=0.5$ (a),
$\nu=0.7$ (b) and $\nu=0.9$ (c). As in the cases of negative on-site
interactions, we may identify the existence of two polaron branches
in the spectrum. However, these two branches seem to behave very differently
from the negative-$U$ case.

First, the low-lying polaron branch always has a notable spectral
width and the width increases with increasing filling factor. This
is remarkably different from the case with $U<0$, where the width
of the sharply peaked low-lying attractive Fermi polarons is negligible
as we increase the filling factor $\nu$ above $0.1$. Second, the
high-lying polaron branch behaves like a well-defined $\delta$-function
peak near the $M$ point, regardless of the filling factor. Finally,
at large filling factor, i.e., $\nu=0.9$, the two branches tend to
connect with each other.

Therefore, in comparison with the negative-$U$ case, the low-lying
and high-lying polaron branches seem to exchange their roles: the
low-lying branch behaves more or less like a repulsive Fermi polaron;
instead the high-lying branch looks like an attractive Fermi polaron,
although it is now restricted to the vicinity of the $M$ point. This
exchange in role becomes much more evident when we increase the on-site
repulsion. In Fig. \ref{fig:fig5_akw2dU12p}, we show the impurity
spectral function at $U=+12t$ and at the filling factor $\nu=0.5$.
The two polaron branches, both in the one-dimensional plot (a) and
in the 2D contour plot (b), are now clearly separated. In particular,
the sharply peaked high-lying polaron branch extends from the $M$-point
to the $\Gamma$-point. Moreover, the high-lying branch at the $X$-point
also becomes well-defined.

\begin{figure}
\begin{centering}
\includegraphics[width=0.5\textwidth]{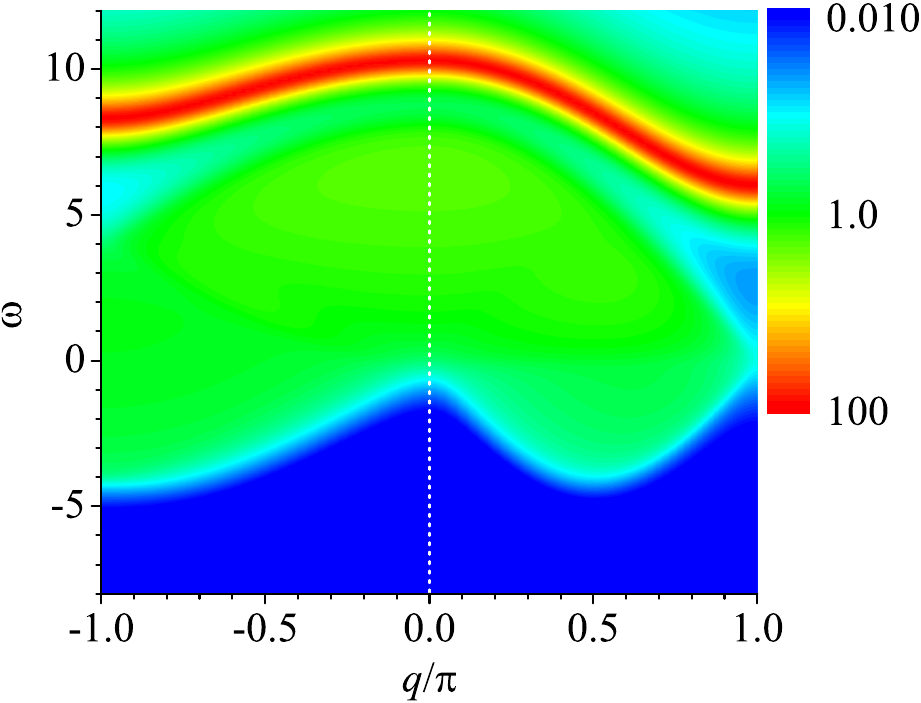}
\par\end{centering}
\caption{\label{fig:fig6_gkw2DU12p} The molecular spectral function $A_{\textrm{mol}}(\mathbf{q},\omega)$
at a large positive interaction strength $U=12t$, in arbitrary units
(as indicated by the color map in the logarithmic scale). For $q<0$,
the wavevector $\mathbf{q}=(q,0)$, while for $q>0$, we consider
the wavevector $\mathbf{q}=(q,q)$ along the diagonal direction. Here,
we take the same temperature $k_{B}T=0.1t$ and filling factor $\nu=0.5$
as in Fig. \ref{fig:fig5_akw2dU12p}.}
\end{figure}

The role exchange is mostly easily understood by considering a particle-hole
transformation for fermionic atoms. At large filling factor above
the half-filling, i.e., $\nu\geq0.5$, it is more convenient to adopt
a viewpoint of holes. We treat unoccupied single-particle states as
holes and introduce the hole creation field operator $h_{\mathbf{k}}^{\dagger}=c_{-\mathbf{k}}$.
When it acts on a fully occupied Fermi sea with unity filling factor
$\nu=1$ (i.e., the vacuum state of holes), it destroys a fermionic
atom with momentum $-\mathbf{k}$ and creates a hole with momentum
$\mathbf{k}$. In the hole representation, the interaction Hamiltonian
in Eq. (\ref{eq:ModelHamiltonian}) can be casted into,
\begin{equation}
\mathcal{H}_{\textrm{int}}=U\sum_{\mathbf{k}}d_{\mathbf{k}}^{\dagger}d_{\mathbf{k}}-\frac{U}{A}\sum_{\mathbf{kk'q}}h_{\mathbf{k}}^{\dagger}d_{\mathbf{q}-\mathbf{k}}^{\dagger}d_{\mathbf{q}-\mathbf{k'}}h_{\mathbf{k'}}.
\end{equation}
Thus, the impurity up-shifts its dispersion relation by an amount
$U$ due to the (mean-field) repulsion of the fully occupied Fermi
sea, and more importantly, the effective interaction between the impurity
and holes becomes attractive, i.e., $U_{\textrm{eff}}=-U<0$. It is
reasonable to assume that this effective attraction would induce attractive
Fermi polarons. As the holes occupy around the $M$ point with a smaller
hole Fermi sea, the density fluctuation - in the form of particle-hole
excitations of the new hole Fermi sea - will first create attractive
Fermi polarons around the $M$ point and then extends to the $\Gamma$
point. 

These attractive polarons are highly non-trivial, in the sense that
they are the highest in energy but remain completely undamped at zero
temperature. It would be useful to name such high-lying Fermi polarons
as super Fermi polarons, to highlight the fact that they are \emph{exact}
many-body states of the system. In contrast, the usual excited Fermi
polaron state, such as repulsive Fermi polaron in the negative-$U$
case, consists of a bundle of many-body states and hence has an intrinsic
decay rate even at zero temperature \cite{Hu2023AB,Wang2024}. We
note that, a similar terminology has been used to characterize the
highest excited many-body state, i.e., super-Tonk-Girardeau state,
in a strongly attractive Bose gas \cite{Astrakharchik2005}.

\begin{figure}
\begin{centering}
\includegraphics[width=0.45\textwidth]{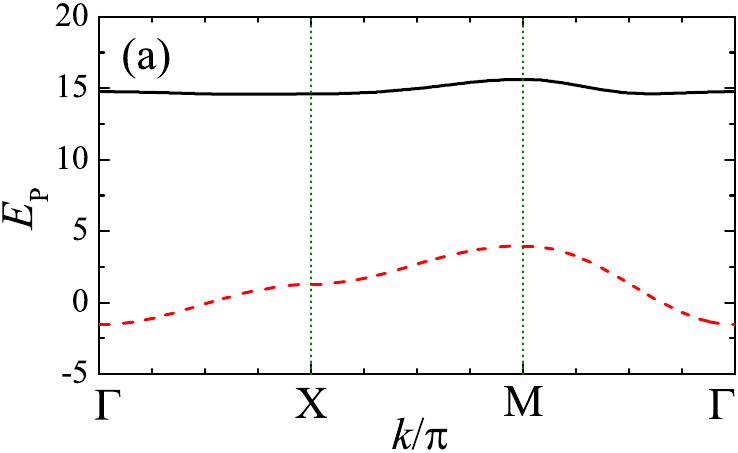}
\par\end{centering}
\begin{centering}
\includegraphics[width=0.45\textwidth]{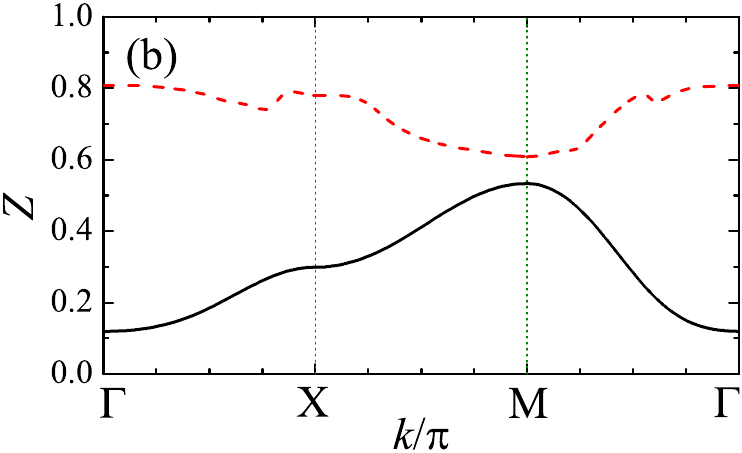}
\par\end{centering}
\begin{centering}
\includegraphics[width=0.45\textwidth]{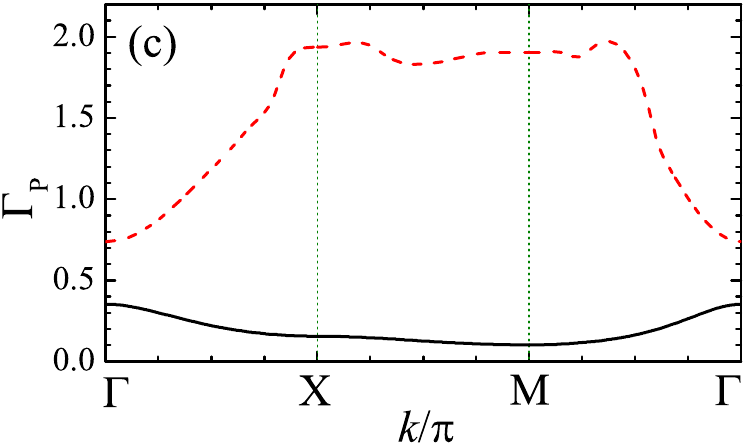}
\par\end{centering}
\caption{\label{fig:fig7_quasiparticleU12p} Quasiparticle properties of Fermi
polarons at the positive interaction strength $U=12t$: polaron energy
(a), residue (b) and decay rate (c). Both polaron energy and decay
rate are measured in units of $t$. The black solid lines show the
results of the upper branch, super Fermi polarons. The red dashed
lines show the properties of the lower branch of standard Fermi polarons.
Along the $\Gamma X$ line and the $XM$ line, we individually increase
$k_{x}$ and $k_{y}$ from 0 to $\pi$, respectively; while along
the $M\Gamma$ line, we decrease both $k_{x}$ and $k_{y}$ from $\pi$
to zero. Once again, we use the tmeperature $k_{B}T=0.1t$ and filling
factor $\nu=0.5$ as in Fig. \ref{fig:fig5_akw2dU12p}.}
\end{figure}

The effective attraction between the impurity and holes may also lead
to a two-body bound state, i.e., a \emph{repulsively} bound pair between
the impurity and fermionic atoms due to repulsion. Actually, such
a repulsion-induced bound pair has already been experimentally observed
in a Bose gas in optical lattices \cite{Winkler2006}. To confirm
the (repulsively) bound pair of the impurity and holes due to the
effective attraction $U_{\textrm{eff}}$, we show in Fig. \ref{fig:fig6_gkw2DU12p}
the molecule spectral function along the $\Gamma M$ line (see the
right part of the figure) and the $\Gamma X$ line (the left part),
at the same parameters as in Fig. \ref{fig:fig5_akw2dU12p}. We see
clearly the molecule peak above the two-particle scattering continuum.
Analogous to the negative-$U$ case, where a two-body bound state
implies the existence of repulsive Fermi polarons, it is natural to
classify the low-lying polaron branch in Fig. \ref{fig:fig5_akw2dU12p}
as repulsive Fermi polarons. In this way, it is not a surprise to
find a nonzero decay rate of low-lying polaron branch, even at temperatures
close to the zero temperature (i.e., $k_{B}T=0.1t$). The decay rate
or spectral broadening of the repulsive Fermi polaron is due to the
scattering with fermionic atoms or holes, since the repulsive polaron
energy is within the two-particle scattering continuum, although the
repulsive polaron turns out to be the low-energy, ground-state-like
polaron quasiparticle.

For completeness, in Fig. \ref{fig:fig7_quasiparticleU12p} we report
the polaron energy (a), residue (b) and decay rate (c) of both polaron
branches along the $\Gamma-X-M-\Gamma$ cut lines, at the same parameters
as in Fig. \ref{fig:fig5_akw2dU12p}. The results of super Fermi polarons
and of repulsive Fermi polarons are shown by the solid lines and dashed
lines, respectively. We find that the dispersion relation of super
Fermi polarons is rather flat, compared with that of repulsive Fermi
polarons, indicating a large effective mass. In particular, the effective
mass of super Fermi polarons at the $M$ point is negative. This is
easy to understand, if we recall the fact that the original mass of
the impurity at the $M$ point (i.e., the top of its energy band)
is negative. On the other hand, for the parameters we choose, super
Fermi polarons always have less spectral weight than repulsive Ferm
polarons, as we infer from the polaron residue. Their weights are
only comparable at the $M$ point, where super Fermi polarons seem
to have the strongest presence. Finally, as we already emphasized,
super Fermi polarons have the smallest decay rate at the $M$ point
only due to thermal broadening. Repulsive Fermi polarons instead always
show a larger decay rate, even at the $\Gamma$ point, where it is
supposed to be most stable. 

\begin{figure}
\begin{centering}
\includegraphics[width=0.5\textwidth]{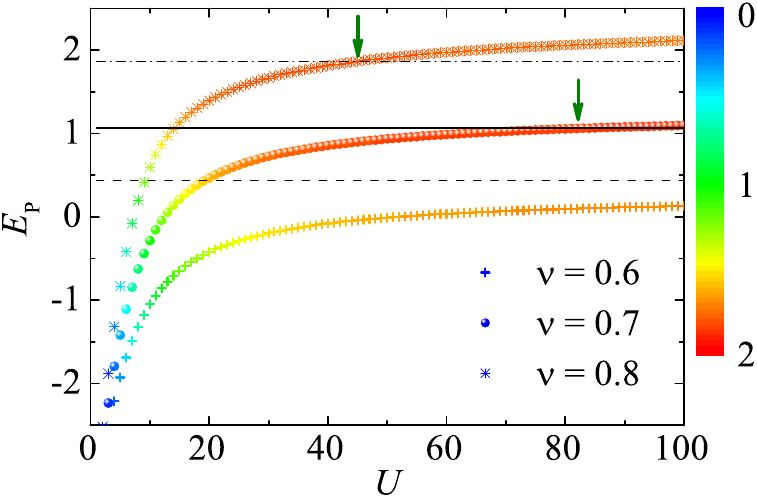}
\par\end{centering}
\caption{\label{fig:fig8_NagaokaFM} Zero-momentum polaron energy $\mathcal{E}_{P}(\mathbf{k}=\mathbf{0})$
as a function of the positive interaction strength $U$, with increasing
filling factor: $\nu=0.6$ (plus symbols), $\nu=0.7$ (circles) and
$\nu=0.8$ (stars). The color of symbols represents the polaron decay
rate $\Gamma_{P}$ in units of $t$, as indicated by the color bar.
From bottom to top, the three horizontal lines show the chemical potential
$\mu$ of the Fermi sea at $\nu=0.6$, $\nu=0.7$ and $\nu=0.8$,
respectively. The two green arrows indicate the critical interaction
strength for Nagaoka ferromagnetism, $U_{c}\simeq84t$ at $\nu=0.7$
and $U_{c}\simeq45t$ at $\nu=0.8$, at which $\mathcal{E}_{P}(\mathbf{0})=\mu$.
Here, we take the temperature $k_{B}T=0.1t$.}
\end{figure}

\subsection{Nagaoka ferromagnetism}

At large on-site repulsion, the polaron problem under investigation
could be related to the celebrated Nagaoka ferromagnetism in a cluster
of spin-1/2 fermions \cite{Nagaoka1966}, which concerns the instability
of a ferromagnetic state with full spin polarization against a single
spin flip. Previous variational studies suggest the breakdown of Nagaoka
ferromagnetism below a certain critical fermion filling factor $\nu_{c}$
or above a corresponding critical hole filling factor $\delta_{c}=1-\nu_{c}$
\cite{Shastry1990,Basile1990,vonderLinden1991,Cui2010}. At infinitely
large repulsion $U=+\infty$, the simple Chevy ansatz predicts $\nu_{c}\simeq0.59$
or $\delta_{c}\simeq0.41$ at zero temperature \cite{Basile1990,Cui2010}.

We may treat the impurity as a single spin-down fermions and all the
others atoms in the Fermi sea as the spin-up fermions. Thus, at large
on-site repulsion $U$ and at large filling factor $\nu>\nu_{c}(U)$,
we may anticipate a phase transition towards the Nagaoka ferromagnetic
state, if we allow the impurity to flip its imaginable spin and to
jump from the zero momentum spin-down state to a single-particle spin-up
state with a momentum $\mathbf{k}\sim\mathbf{k}_{F}\sim(\pm\pi,\pm\pi)$.
Accordingly, the Fermi sea will shuffle its Fermi surface to satisfy
the momentum conservation. This anticipation reasonably agrees with
the filling factor $\nu$-dependence of both Fermi polaron branches
at large repulsion $U\gg t$, as we observe in Fig. \ref{fig:fig4_akw2dU6p}.
As the filling factor increases, the tendency of the spin reversal
makes low-lying repulsive Fermi polarons less well-defined and at
the same time makes high-lying super Fermi polarons much more sharply
peaked. Thus, in the thermodynamic limit, upon infinitesimal fluctuations
in temperature and lattice potential, the fragile low-lying repulsive
polaron state can easily turn into a state, where a super Fermi polaron
with momentum $\mathbf{k}\sim\mathbf{k}_{F}$ gains notable weight.
As the super Fermi polaron might be viewed as the Nagaoka ferromagnetic
state after the imaginable spin-flip, there is a thermodynamic instability
of turning the low-lying repulsive polaron state into the Nagaoka
ferromagnetic state, if the spin-reversal is allowed.

As a quantitative measure, we may consider the lowest energy of the
repulsive Fermi polaron at the $\Gamma$ point, and compare it to
the chemical potential $\mu$, which can be regarded as the energy
of the spin-up state after the imaginable spin flip. The stability
of the Nagaoka ferromagnetic state is then ensured by the condition,
$\mathcal{E}_{P}(0)>\mu$. Although the repulsive Fermi polaron is
not a single quantum many-body state, we believe that this condition
could provide a reasonable thermodynamic evaluation of the critical
on-site repulsion at a given filling factor, $U_{c}(\nu)$, at very
low temperature. 

In Fig. \ref{fig:fig8_NagaokaFM}, we compare the low-lying repulsive
polaron energy $\mathcal{E}_{P}(0)$ with the chemical potential $\mu$
with increasing on-site repulsion, at three filling factors as indicated.
At $\nu=0.6$, we always find that the polaron energy is below the
chemical potential, indicating the absence of the Nagaoka ferromagnetic
state at the on-site repulsion considered in the figure. This is understandable,
since $\nu=0.6$ is very close to the critical filling factor $\nu_{c}\simeq0.59$
at $U=+\infty$. The small but non-zero temperature $k_{B}T=0.1t$
used in our calculations effectively reduce the interaction effect
and may already wash out the Nagaoka ferromagnetism transition. In
contrast, at other two filling factors in the figure, by using the
criterion $\mathcal{E}_{P}(0)=\mu$ we find $(W/U)_{c}\simeq0.10$
at $\nu=0.7$ and $(W/U)_{c}\simeq0.18$ at $\nu=0.8$, where $W=8t$
is the energy band width of the square lattice. These two critical
values $(W/U)_{c}$ agree qualitatively well with the initial estimation
by Shastry, Krishnamurthy and Anderson \cite{Shastry1990}, and the
improved variational result by von der Linden and Edwards \cite{vonderLinden1991}. 

\subsection{Finite temperature effect}

\begin{figure}
\begin{centering}
\includegraphics[width=0.25\textwidth]{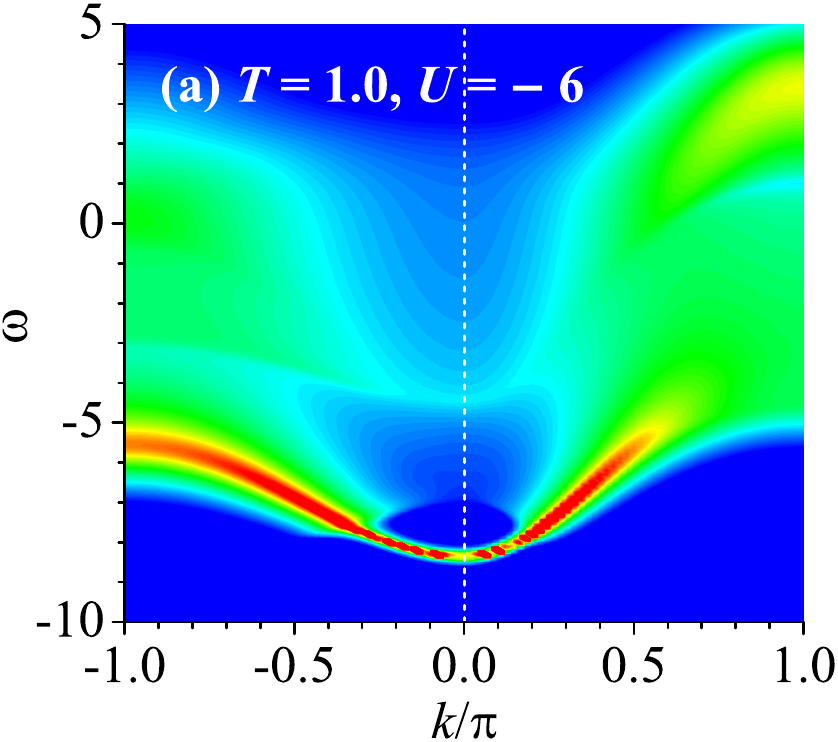}\includegraphics[width=0.25\textwidth]{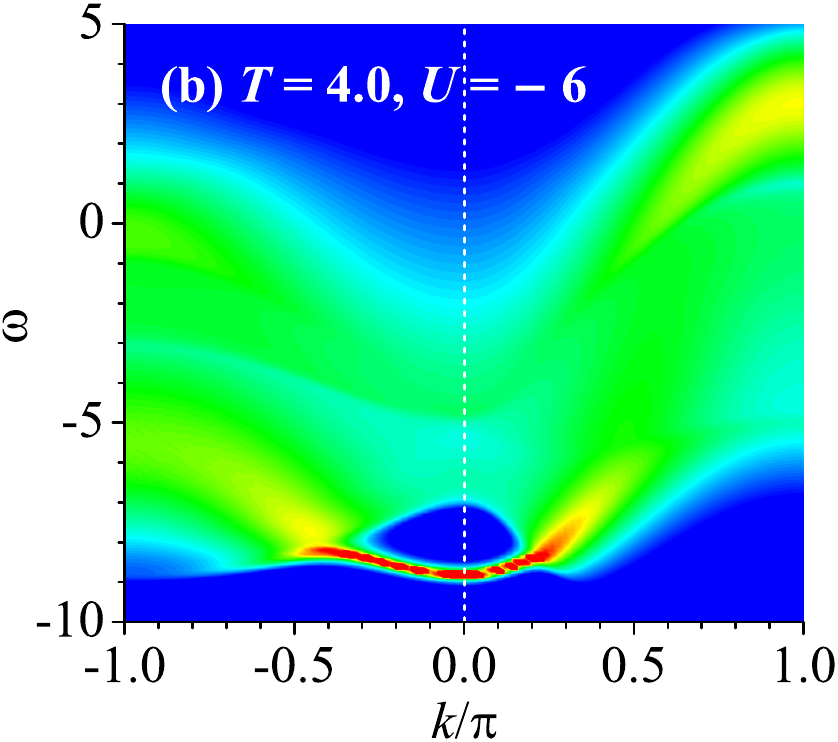}
\par\end{centering}
\begin{centering}
\includegraphics[width=0.25\textwidth]{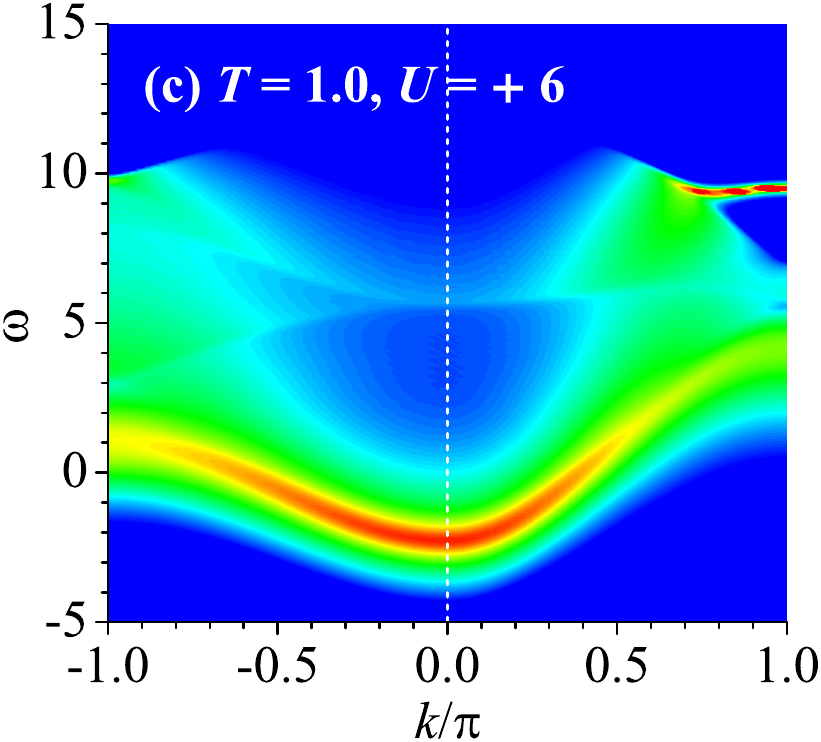}\includegraphics[width=0.25\textwidth]{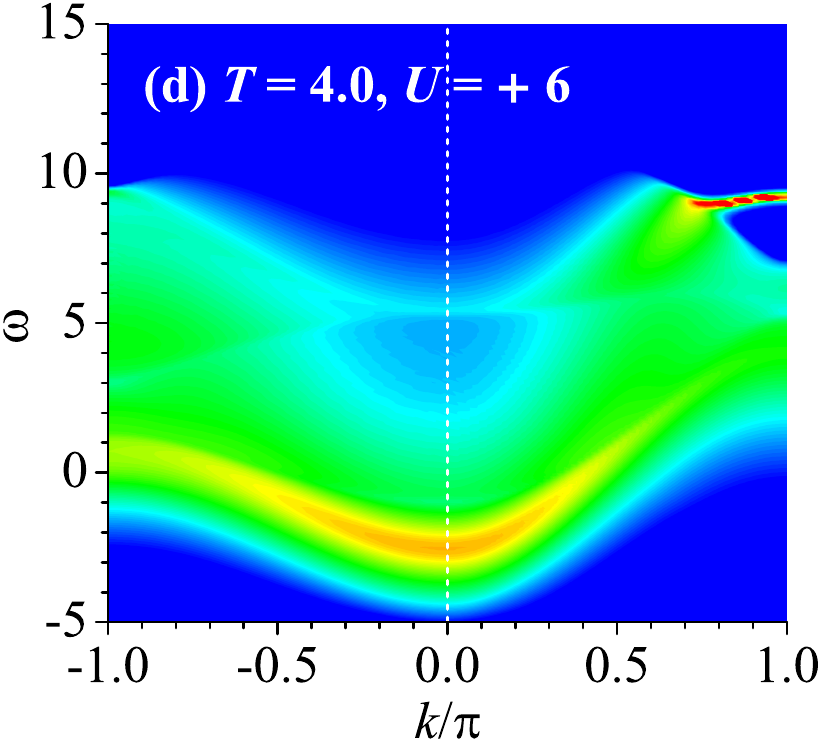}
\par\end{centering}
\caption{\label{fig:fig9_tdep} Impurity spectral function $A(\mathbf{k},\omega)$
at large temperatures $k_{B}T=t$ (a and c) and $k_{B}T=4t$ (b and
d). The upper panel and low panel report the results at $U=-6t$ and
$U=+6t$, respectively. We use the same logarithmic contour plots
as in Fig. \ref{fig:fig1_akw2dU6m} and consider the half-filling
of the Fermi sea, $\nu=0.5$.}
\end{figure}

We finally briefly discuss the temperature effect. In Fig. \ref{fig:fig9_tdep},
we show the 2D contour plot of spectral function at the temperature
$T=t$ (see the two subplots (a) and (c) on the left) and $T=4t$
(the right two subplots (b) and (d)). We focus on a filling factor
$\nu=0.5$ and consider both on-site attractions $U=-6t$ (see the
upper panel) and repulsions $U=+6t$ (the low panel). We observe that
the conventional attractive Fermi polarons in the negative-$U$ case
significantly changes with increasing temperature. Quite differently,
super Fermi polarons near the $M$ point with on-site repulsions appear
to be insensitive to temperature and hence are thermally robust. They
remains sharply peaked at the temperature as large as the half energy
band width (i.e., $k_{B}T=4t=W/2$). 

\section{Conclusions and outlooks}

In conclusion, we have investigated the Fermi polaron problem in two-dimensional
square lattices at finite temperature, with both on-site attractive
interactions and repulsive interactions between an impurity and a
Fermi sea of non-interacting fermions. The standard non-self-consistent
many-body $T$-matrix approach has been used, which well describes
the key ingredient of polaron physics, i.e., the one-particle-hole
excitations of the Fermi sea as excited by the on-site interaction
\cite{Combescot2007,Hu2022a}. This method is equivalent to a variational
ansatz previously used to address Nagaoka ferromagnetism \cite{Nagaoka1966}
of a cluster of spin-1/2 fermions on square lattices \cite{Shastry1990,Basile1990,vonderLinden1991,Cui2010}.
However, our diagrammatic calculations are able to obtain the impurity
spectral function at finite temperature, thereby leading to new understanding
to the old research topic of Nagaoka ferromagnetism.

For on-site attractions at small filling factor, we have found conventional
Fermi polarons, including both attractive and repulsive branches.
In the dilute limit of vanishingly small filling factor, the results
can be well understood by using a free-space Fermi polaron model \cite{Parish2011,Parish2013,Bour2015}.
We have demonstrated how the polaron physics is affected by the lattice
structure. In particular, we have shown that the repulsive Fermi polaron
at the $M$ point, where $\boldsymbol{\mathbf{k}}_{M}=(\pm\pi,\pm\pi)$,
is relatively robust with increasing filling factor, due to the stable
two-body bound state at the corner of the first Brillouin zone.

For on-site repulsions at large filling factor, we have found a novel
type of Fermi polarons, the so-called super Fermi polaron, when the
repulsion is strong enough. The super Fermi polaron is an exact many-body
state centered around the $M$ point and is therefore long-lived at
low temperature, although it is highly excited with large energy.
We have explained that the formation mechanism of high-lying super
Fermi polarons is due to an effective attraction between the impurity
and holes arising from strong on-site repulsions. Therefore, it can
be understood in terms of conventional attractive Fermi polarons.
We have shown that there is also a ground-state-like, low-lying Fermi
polaron branch with on-site repulsion. However, this low-lying polaron
branch has a finite decay rate and should be understood as conventional
repulsive polarons. 

The classification of the two polaron branches in the case of on-site
repulsions suggests that the appearance of the super Fermi polaron
could be viewed as a precursor of Nagaoka ferromagnetism. This is
because, at large filling factor with increasing on-site repulsions,
the impurity may leave from the short-lived repulsive polaron state
at zero momentum, virtually occupy the much more well-defined super
Fermi polaron state at the $M$ point, and turn the system into the
Nagaoka ferromagnetic state upon reversing its imaginable spin. We
have provided a thermodynamic estimation for the critical on-site
repulsion $U_{c}$ needed for the transition into a Nagaoka ferromagnetic
state, at a given large filling factor $\nu\sim1$. The obtained values
of $U_{c}(\nu)$ agree qualitatively well with the previous variational
calculations \cite{Shastry1990,vonderLinden1991}.

In future studies, it would be useful to improve theoretical predictions
on super Fermi polarons beyond the non-self-consistent many-body $T$-matrix
approximation. This would provide us an accurate determination of
the phase diagram for the Nagaoka ferromagnetic phase transition,
at both zero temperature and finite temperature. It would also motivate
the experimental investigation of the intriguing Nagaoka ferromagnetism
in cold-atom laboratories, by preparing a spin-population imbalanced
Fermi gas in two-dimensional optical lattices \cite{Mazurenko2017}.
\begin{acknowledgments}
This research was supported by the Australian Research Council's (ARC)
Discovery Program, Grants Nos. DP240101590 (H.H.), FT230100229 (J.W.),
and DP240100248 (X.-J.L.).
\end{acknowledgments}

\appendix
\begin{figure}
\begin{centering}
\includegraphics[width=0.5\textwidth]{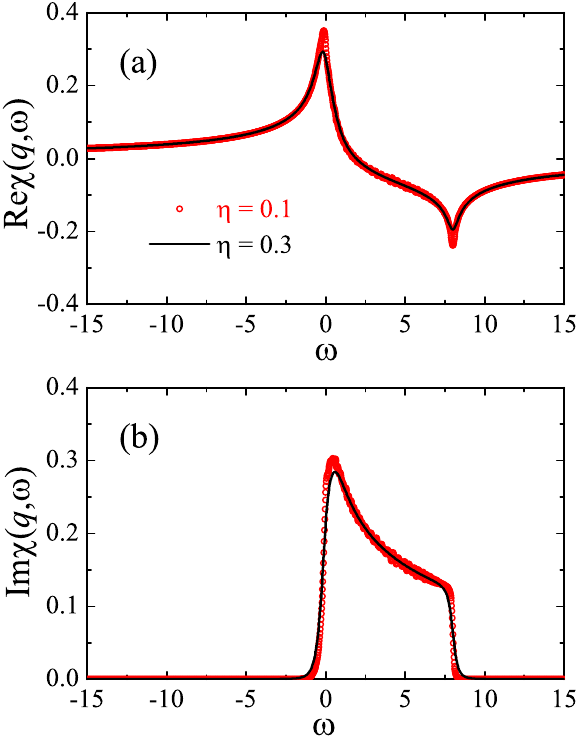}
\par\end{centering}
\caption{\label{fig:figA1_kappa} The real part and imaginary part of the pair
propagator $\chi(\mathbf{q},\omega)$ at zero wavevector $\mathbf{q}=\mathbf{0}$,
in arbitrary units. The black lines and red circles show the results
with $\eta=0.3t$ and $\eta=0.1t$, respectively. The temperature
is set to $k_{B}T=0.1t$.}
\end{figure}

\section{The choice of the broadening factor}

In Fig. \ref{fig:figA1_kappa}, we report the pair propagator 
\begin{equation}
\chi\left(\mathbf{q},\omega\right)=\Gamma^{-1}\left(\mathbf{q},\omega\right)-\frac{1}{U}
\end{equation}
calculated at two broadening factors, $\eta=0.1t$ (circles) and $\eta=0.3t$
(lines), following the linear extrapolation scheme in Eq. (\ref{eq:LinearExtrapolation}).
We find the results of $\chi(\mathbf{q},\omega)$ are independent
on $\eta$, except at the frequencies $\omega\sim0$ and $\omega\sim8t$,
where its real part exhibits sharp peaks and its imaginary part starts
to appear or disappear. At the small broadening factor $\eta=0.1t$,
the insufficient number of grid points used in our gaussian quadrature
integration leads to a small oscillation in the calculated pair propagator.
This unwanted oscillatory behavior can be quickly removed by increasing
$\eta$ to $0.3t$. A nonlinear extrapolation can also be implemented
to improve the numerical accuracy, but it might not be necessary,
considering our purpose of clarifying the existence of super Fermi
polarons.

\end{document}